\documentclass[12pt]{article}

\usepackage{amsmath,amssymb,amsfonts}
\usepackage{array}
\usepackage{algorithm2e}
\usepackage{tabularx}
\usepackage{subcaption}
\usepackage{graphicx}
\usepackage[biblabel]{cite}
\usepackage[colorlinks=true, linkcolor=blue, citecolor=blue, urlcolor=blue]{hyperref}

\usepackage[a4paper,margin=25mm]{geometry}

\begin{document}

\title{USDs: A universal stabilizer decoder framework using symmetry}

\author{
    Hoshitaro Ohnishi\thanks{Computer Science Program, Graduate School of Science and Technology, Meiji University, 1-1-1 Higashimita, Tama-ku, Kawasaki, Kanagawa 214-8571, Japan}
    \and
    Hideo Mukai\thanks{Computer Science Program, Graduate School of Science and Technology, Meiji University, 1-1-1 Higashimita, Tama-ku, Kawasaki, Kanagawa 214-8571, Japan; Department of Computer Science, School of Science and Technology, Meiji University, 1-1-1 Higashimita, Tama-ku, Kawasaki, Kanagawa 214-8571, Japan. Corresponding Author, e-mail: mukai@meiji.ac.jp} 
}

\date{}

\maketitle

\begin{abstract}
Quantum error correction is indispensable to achieving reliable quantum computation. When quantum information is encoded redundantly, a larger Hilbert space is constructed using multiple physical qubits, and the computation is performed within a designated subspace.

When applying deep learning to the decoding of quantum error-correcting codes, a key challenge arises from the non-uniqueness between the syndrome measurements provided to the decoder and the corresponding error patterns that constitute the ground-truth labels. Building upon prior work that addressed this issue for the toric code by re-optimizing the decoder with respect to the symmetry inherent in the parity-check structure, we generalize this approach to arbitrary stabilizer codes.

In our experiments, we employed multilayer perceptrons to approximate continuous functions that complement the syndrome measurements of the Color code and the Golay code. Using these models, we performed decoder re-optimization for each code. For the Color code, we achieved an improvement of approximately 0.8\% in decoding accuracy at a physical error rate of 5\%, while for the Golay code the accuracy increased by about 0.1\%. Furthermore, from the evaluation of the geometric and algebraic structures in the continuous function approximation for each code, we showed that the design of generalized continuous functions is advantageous for learning the geometric structure inherent in the code. Our results also indicate that approximations that faithfully reproduce the code structure can have a significant impact on the effectiveness of reoptimization.

This study demonstrates that the re-optimization technique previously shown to be effective for the Toric code can be generalized to address the challenge of label degeneracy that arises when applying deep learning to the decoding of stabilizer codes.

\end{abstract}

\section{Introduction}
Quantum computation has the potential to outperform classical computation for certain classes of problems by exploiting the principle of quantum superposition, which enables computational processes that are impossible for classical systems \cite{preskill2018, arute2019}. However, the qubits that constitute the fundamental units of quantum information are highly susceptible to external noise, making the execution of reliable computations significantly more challenging than in classical computing \cite{nielsen2000, terhal2013}. 

To mitigate the impact of external noise on qubits, quantum error correction techniques are indispensable. In fault-tolerant quantum computation, logical operations are performed within a designated subspace embedded in a larger Hilbert space constructed from multiple physical qubits. The regions outside this computational subspace provide redundancy that enables errors to be detected. When the state of the physical qubits deviates from the code subspace due to noise, such deviations are detected through syndrome measurements obtained by evaluating the stabilizer generators.

The decoding task for a quantum error-correcting code can thus be viewed as the problem of inferring the underlying error pattern acting on the qubits from the observed syndrome outcomes. For small codes, such as the Steane or Shor codes, lookup-table (LUT) decoding may be feasible \cite{steane1996, reichardt2018, shor1995, calderbank1995}. However, for large-scale codes, the number of possible syndrome patterns becomes prohibitively large, necessitating the development of deterministic algorithms such as minimum-weight perfect matching (MWPM), or the use of machine-learning-based decoders \cite{fowler2018}. 

Surface code represents a highly promising paradigm in quantum error correction, and numerous decoding approaches based on machine learning have been proposed. For the Toric code, which is one specific realization of surface code, decoders based on a variety of neural network architectures including multilayer perceptrons (MLPs), convolutional neural networks (CNNs), graph neural networks (GNNs), and Transformers have been developed \cite{kitaev2003a, krastanov2017a, wang2022a, fjelddahl2024a}. 

However, a common issue arises in the training of these decoders: the syndrome measurements used as input do not uniquely correspond to the error patterns that serve as the ground-truth labels. To address this problem, Ohnishi proposed a method in which a continuous function that mathematically complements the Toric code’s syndrome measurements is approximated using a multilayer perceptron, and the resulting model is then used to re-optimize the decoder \cite{symmetry2025}. This approach enables the decoder to output correction operators that respect the symmetry of the parity-check structure, leading to observable improvements in the decoding performance for the Toric code. 

We extend this method by reformulating it so that it can be applied to arbitrary stabilizer codes. By defining a continuous function that mathematically complements the computational process in which eigenvalue measurements are performed on the qubit states using the stabilizer generators of the target code, our framework enables decoder re-optimization that is independent of the specific structure of the code or the number of qubits required for its encoding. 

In our experiments, we applied this method to a Transformer decoder composed solely of an encoder block pretrained to perform decoding for the Color code and Golay code. At a physical error rate of 5\%, we observed an improvement of approximately 0.8\% in decoding accuracy for the Color code and about 0.1\% for the Golay code. These results demonstrate that re-optimization based on the symmetry of the parity checks implemented through a continuous function defined solely from the structure of the stabilizer generators can yield measurable performance gains. Furthermore, the results of the structural evaluation of continuous function approximations for each code indicate that the design of generalized continuous functions is more effective at reproducing the geometric structure of a code than its algebraic structure. In particular, we observed a larger reoptimization effect for the color code, which possesses pronounced geometric features, than for the Golay code, which is characterized by strong algebraic symmetry. These findings suggest that, in continuous function approximation, the extent to which the structural characteristics of a code can be faithfully reproduced may play a crucial role in achieving effective reoptimization.

\section{Method} 
\label{2}
In this study, we first define a continuous function that mathematically complements the syndrome measurements of general stabilizer codes in a manner independent of their geometric or algebraic structure, as well as of the number of qubits required for encoding. This continuous function is then approximated using a multilayer perceptron. Next, for both the Color code \cite{bombin2006} and the Golay code \cite{calderbank1997}, we train a Transformer to learn the mapping between syndrome measurements and their corresponding error patterns. Finally, we re-optimize the pretrained Transformer decoder using the model that approximates the continuous function and compare the decoding accuracy before and after applying the re-optimization procedure. The Transformer architecture consists solely of encoder blocks, and the decoding task for each code is formulated as a regression problem. 

\subsection{Problem Setting} 
\label{2.1}
\begin{figure}[t]
\centering
\includegraphics[width=0.35\linewidth]{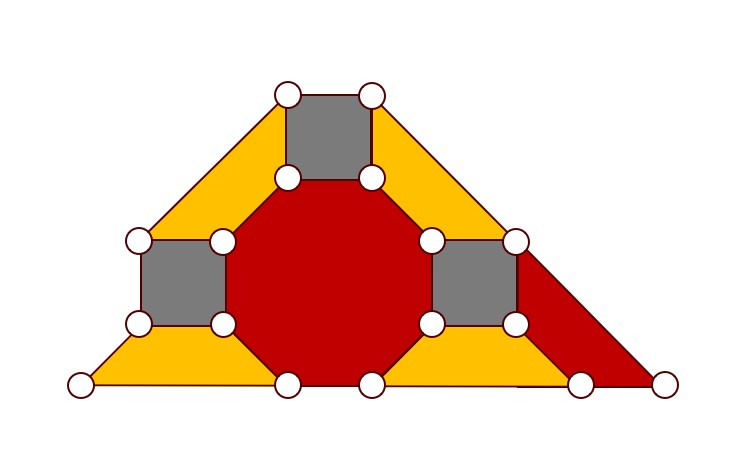}
\caption{Conceptual illustration of the color code with code distance 5. White circles represent data qubits, and each face corresponds to a stabilizer generator acting on the adjacent data qubits.}
 \label{fig:color_code}
\end{figure}

In this study, we consider the (4,8,8) Color code with code distance 5, as well as the [[23,1,7]]Golay code. A schematic illustration of the color code is shown in Fig.~\ref{fig:color_code}.

In addition, the Golay code is defined by the following generator polynomials, each of weight 8.
\begin{equation}
 h_1(x) = x^{12} + x^{10} + x^{7} + x^4 + x^3 + x^2 + x + 1
\end{equation}

Let $n$ denote the number of qubits required for encoding in each code. Here, we assume that each qubit may independently experience both an X-error and a Z-error. Under this assumption, the total number of possible error patterns is $2^{2n}$. On the other hand, the number of stabilizer generators is $n-1$, and each stabilizer yields an eigenvalue of either +1 or -1upon measurement. Therefore, the number of possible syndrome patterns corresponding to errors is $2^{n-1}$. 

In our formulation, we represent the error pattern by concatenating two binary vectors that separately indicate the presence (1) or absence (0) of X-errors and Z-errors on each qubit and denote the resulting vector by $E \in {0,1}^{2n}$. Similarly, the outcome of the syndrome measurement is represented by mapping eigenvalues +1 to 0 and -1 to 1, yielding a binary vector $M \in {0,1}^{n-1}$. In order to apply machine learning to decoding of a quantum error-correcting code, the task can thus be formulated as the following regression problem.
\begin{equation}
 [0,1]^{n-1} \to [0,1]^{2n}
\end{equation}

We assume that the probability distribution governing the errors acting on the qubits follows a discrete uniform model. Each qubit may experience an X-error, a Z-error, or a Y-error, which includes the simultaneous occurrence of both X- and Z-errors. Denoting their occurrence probabilities by $p_x$, $p_z$, and $p_y$, respectively, we introduce a single parameter $p$ and set $p_x = p_z = p_y = \frac{p}{3}$.

The decoder performance is evaluated by simulating 10,000 decoding trials for each value of $p$, where the error occurrences follow the probability distribution defined above and $p$ varies from 0.001 to 0.05 in increments of 0.0001. A decoding attempt is deemed successful if, after applying the correction, all syndromes are resolved and no logical operator has been inadvertently applied to the code state. Any unsuccessful outcome is regarded as a logical error, and the decoder is assessed by computing the resulting logical error rate. 

\subsection{Continuous Extension of Syndrome Measurements Applicable to Arbitrary Stabilizer Codes} 
\label{2.2}
In stabilizer codes employing $n$ qubits for encoding, syndrome measurements are performed by evaluating the eigenvalues of the $n-1$ stabilizer generators. Using the notation introduced earlier, this process can be regarded as a mapping from the error configuration represented as an element of $\{0,1\}^{2n}$ to a sequence of eigenvalues, represented as one of the $2^{n-1}$ possible syndrome patterns. This correspondence is fully determined by the stabilizer generators that define the code. Here, for each code, we will approximate a continuous function $[0,1]^{2n} \to [0,1]^{n-1}$ that extends this correspondence by using a multilayer perceptron. The continuous function approximated is expressed as follows. 
\begin{equation}
 f:[0,1]^{2n} \to [0,1]^{n-1}
\end{equation}
\begin{equation}
 f(E) = (f_1{E),f_2(E),\dots,f_{n-1}(E)})
\end{equation}
\begin{equation}
 f_i(E) = \frac{1-cos(v_i\pi)}{2}
\end{equation}
\begin{equation}
 v_i = \Sigma_{j=1}^n{E_j S_{i,j}^Z + E_{j+n} S_{i,j}^X}
\end{equation}

Let $S_{i,j}^X$ denote a binary indicator that takes the value 1 if the i-th stabilizer generator applies an X-operation to the j-th qubit, and 0 otherwise. Similarly, let $S_{i,j}^X$ indicate whether the operation is a Z-operation. Under these definitions, the function above returns the same value as the syndrome measurement when given the error-state vector $E$ as input. Because this function is continuous and defined on a compact subset of Euclidean space, the universal approximation theorem ensures that it can be approximated to arbitrary accuracy by a multilayer perceptron \cite{cybenko1989, hornik1989, leshno1993}. 

The input data used for approximating the continuous function with a multilayer perceptron consists of vectors sampled from the domain $[-0.5,1]^{2n}$, where each component is independently drawn from a uniform distribution. The choice of the interval $[-0.5,1]^{2n}$ is made to ensure accurate learning of the output corresponding to the zero vector and to prevent improper approximation that may arise when sampling only on the boundary of the domain. The remaining hyperparameters used in training the model are listed in Table~\ref{tab:ApproxNN}.
\begin{table}[htbp]
\centering
\renewcommand{\arraystretch}{1.0}
\begin{tabular}{@{} c|c @{}}
Parameters & Values \\
\cline{1-2}
Number of hidden layers & $1$ \\
Hidden layer dimension & $1000$ \\
Activation function & SeLU \\
Output layer activation function & Sigmoid \\
Number of training data & $10^7$ \\
Number of test data & $10^5$ \\
Batch size & $1000$ \\
Epoch & $50$ \\
Learning rate & $0.0001$ \\
Loss function & MSE \\
Optimizer & AdamW \\
\end{tabular}
\caption{Neural network training settings for approximating a continuous function.}
\label{tab:ApproxNN}
\end{table}

\subsection{Training and Re-optimization of the Decoder} 
\label{2.3}
We first train a Transformer to learn the correspondence between error patterns and the resulting syndrome measurements for each code. The hyperparameters used during this training stage are listed in Table~\ref{tab:FirstTrain}. Next, we re-optimize the weights of the pretrained Transformer decoder by using a multilayer perceptron that approximates the continuous syndrome function. The dataset used for re-optimization is identical to the one used in the initial training phase.
\begin{table}[htbp]
\centering
\renewcommand{\arraystretch}{1.0}
\begin{tabular}{@{} c|c|c @{}}
Parameters & Color & Golay \\
\cline{1-3}
Number of head & $8$ & $8$ \\
Number of encoder layer & $4$ & $4$ \\
Embedded dimension & $128$ & $128$ \\
Output layer activation function & Sigmoid & Sigmoid \\
Number of training data & $10^6$ & $10^6$ \\
Number of test data & $10^5$ & $10^5$ \\
Batch size & $1000$ & $1000$ \\
Epoch & $50$ & $30$ \\
Learning rate & $0.0001$ $0.0001$ \\
Loss function & BCE &BCE \\
Optimizer & RAdam & RAdam\\
\end{tabular}
\caption{Settings for training Transformer decoder.}
\label{tab:FirstTrain}
\end{table}

In the re-optimization procedure, we first input a syndrome vector into the decoder to obtain a predicted error vector. We then compute the element-wise absolute difference between this prediction and the true error vector. This difference vector is fed into the multilayer perceptron, which produces a pseudo-estimated syndrome vector corresponding to the corrected state. If the final output is the zero vector, it indicates that the correction operator produced by the decoder successfully returns the state to the code space without applying any logical operator. 

Thus, during re-optimization, the weights of the multilayer perceptron are kept fixed, and only the weights of the Transformer decoder are updated so that the final output approaches the zero vector. The hyperparameters used in the re-optimization phase are summarized in Table~\ref{tab:Reoptim}.

\begin{table}[htbp]
\centering
\renewcommand{\arraystretch}{1.0}
\begin{tabular}{@{} c|c|c @{}}
Parameters & Color & Golay \\
\cline{1-3}
Batch size & $1000$ & $1000$ \\
Epoch & $75$ & $75$ \\
Learning rate & $10^{-7}$ & $10^{-7}$ \\
Loss function & BCE &BCE \\
Optimizer & RAdam & RAdam\\
\end{tabular}
\caption{Settings for reoptimization of Transformer decoder.}
\label{tab:Reoptim}
\end{table}

\section{Results and discussion} 
\label{3}
\subsection{Approximation Accuracy of Continuous Functions} 
\label{3.1}
We evaluated the accuracy of the multilayer perceptron that approximates the continuous functions associated with each code by using 10,000 samples drawn from the same probability distribution as the one used during training. As evaluation metrics, cosine similarity, mean squared error (MSE), and mean absolute error (MAE) were employed. Table~\ref{tab:result_app} shows the average values of these metrics. As the table indicates, the continuous functions are approximated with high accuracy. 
\begin{table}[htbp]
\centering
\renewcommand{\arraystretch}{1.0}
\begin{tabular}{@{} c|c|c @{}}
evaluation index& Color & Golay \\
\cline{1-3}
cosine similarity & $0.95835$ & $0.93312$ \\
MSELoss & $0.02415$ & $0.04752$ \\
MAELoss & $0.05373$ & $0.12284$ \\
\end{tabular}
\caption{Results of evaluation of the ability of model using a continuous function approximation.}
\label{tab:result_app}
\end{table}

We further evaluated the extent to which the multilayer perceptron (MLP) used for approximation can reproduce the structural characteristics of each code from both geometric and algebraic perspectives. For the evaluation of geometric structure, we employed the Dirichlet energy $E(\cdot)$, and computed the ratio $\frac{E(f_{mlp})}{E(f)}$ between the original function $f$ and its MLP approximation $f_{mlp}$ \cite{dirichlet2024}. The algebraic structure was evaluated by examining invariance under the action of the stabilizer generators of each code. Specifically, for each generator $s$, we calculated the mean squared error $MSE(f(x), f_{mlp}(sx))$. The results are summarized in Table~\ref{tab:result_app_arc}. From the table, we observe that the geometric structure is more faithfully reproduced for the color code than for the Golay code. In contrast, with respect to algebraic structure, comparison with the values in Table~\ref{tab:result_app} indicates no significant difference from the overall decoding accuracy. 
\begin{table}[htbp]
\centering
\renewcommand{\arraystretch}{1.0}
\begin{tabular}{@{} c|c|c @{}}
evaluation index& Color & Golay \\
\cline{1-3}
Dirichlet energy & $0.88024$ & $0.64704$ \\
group invariance & $0.02415$ & $0.04752$ \\
\end{tabular}
\caption{Results of structural evaluation of continuous function approximation using MLP.}
\label{tab:result_app_arc}
\end{table}

\subsection{Effect of Re-optimization in the Color Code} 
\label{3.2}
The initial training of the decoder was performed until the decrease in loss converged. This was done to ensure that any change in accuracy observed after re-optimization can be attributed solely to the additional effect of the re-optimization process. Decoder training and re-optimization were each carried out 20 times, and the performance of each model was evaluated.

\begin{figure}[htbp]
 \centering
 \begin{subfigure}[t]{0.45\linewidth}
 \centering
 \includegraphics[width=\linewidth]{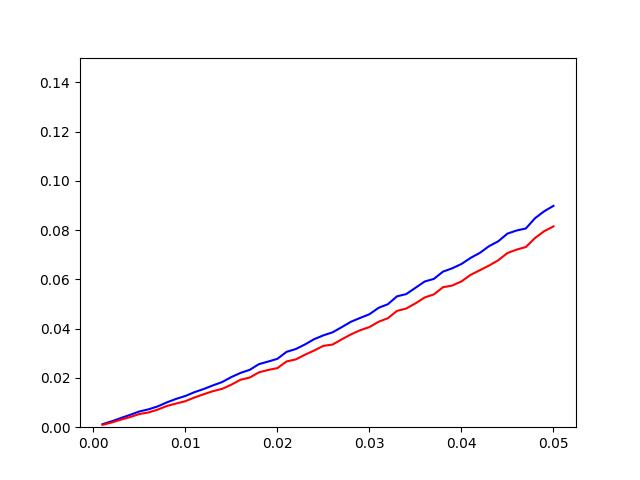}
\caption{The vertical axis represents the logical error rate of the decoder, and the horizontal axis represents the physical error rate. The red curve shows the values before reoptimization, while the blue curve shows the values after reoptimization.}
 \label{fig:color_mean}
 \end{subfigure}
 \hspace{0.05\linewidth}
 \begin{subfigure}[t]{0.45\linewidth}
 \centering
 \includegraphics[width=\linewidth]{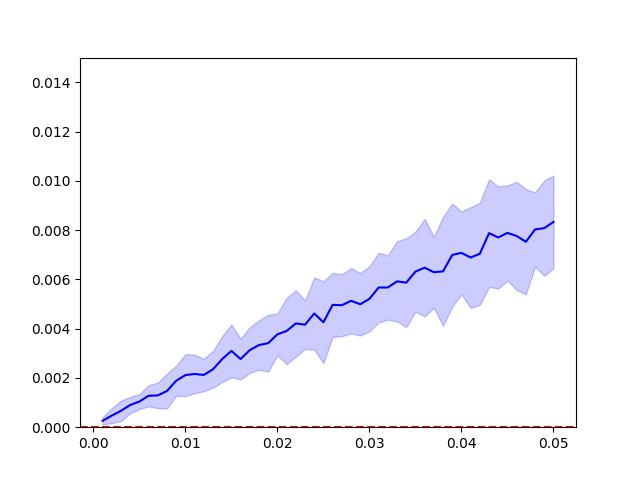}
\caption{The vertical axis represents the difference in the logical error rate before and after reoptimization, and the horizontal axis represents the physical error rate. The solid line indicates the mean value, and the shaded blue region represents the range within one standard deviation from the mean.}
 \label{fig:color_diff}
 \end{subfigure}
\caption{Results of experiments on the Color code.}
 \label{fig:color_result}
\end{figure}

Figure~\ref{fig:color_mean} shows the average logical error rates of the decoder before and after reoptimization, while Fig.~\ref{fig:color_diff} illustrates the average reduction in logical error rate together with the corresponding standard deviation. As the results indicate, our method consistently provides a certain level of accuracy improvement for the Color code. 

\subsection{Effect of Re-optimization in the Golay Code} 
\label{3.3}
As in the case of the Color code, the initial training of the decoder was performed until the decrease in the loss had converged. Training and re-optimization were each conducted 20 times. 

\begin{figure}[htbp]
 \centering
 \begin{subfigure}[t]{0.45\linewidth}
 \centering
 \includegraphics[width=\linewidth]{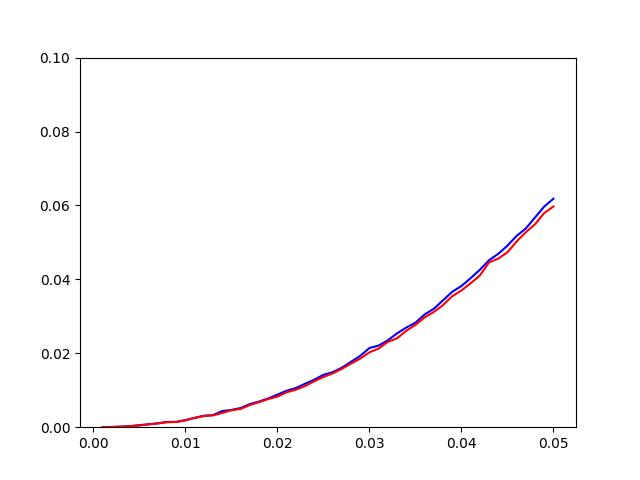}
\caption{The vertical axis represents the logical error rate of the decoder, and the horizontal axis represents the physical error rate. The red curve shows the values before reoptimization, while the blue curve shows the values after reoptimization.}
 \label{fig:golay_mean}
 \end{subfigure}
 \hspace{0.05\linewidth}
 \begin{subfigure}[t]{0.45\linewidth}
 \centering
 \includegraphics[width=\linewidth]{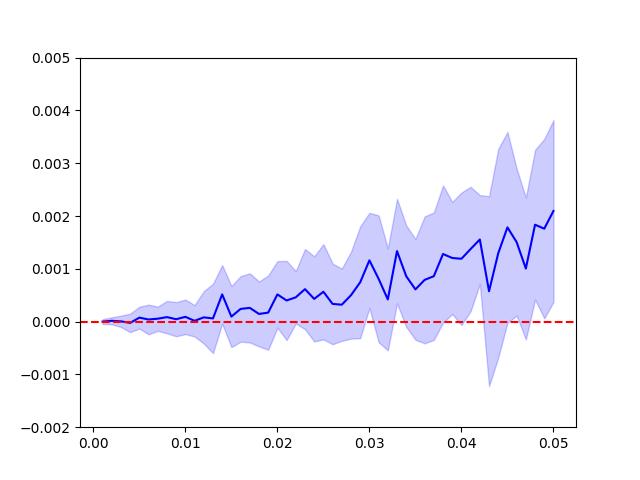}
\caption{The vertical axis represents the difference in the logical error rate before and after reoptimization, and the horizontal axis represents the physical error rate. The solid line indicates the mean value, and the shaded blue region represents the range within one standard deviation from the mean. The red dashed line indicates zero.}
 \label{fig:golay_diff}
 \end{subfigure}
\caption{Results of experiments on the Golay code.}
 \label{fig:golay_result}
\end{figure}

Figure~\ref{fig:golay_mean} presents the average logical error rates of the decoder before and after reoptimization for the Golay code, while Fig.~\ref{fig:golay_diff} shows the average reduction in logical error rate together with its standard deviation. As the results indicate, the effect of re-optimization in the Golay code does not lead to a consistent improvement in accuracy; however, on average, it does provide a positive impact on the decoder's performance.

\subsection{Comparison and Discussion of Re-optimization Effects Across Codes}
\label{3.4}
Our method yielded larger and more stable improvements in decoding accuracy for the Color code than for the Golay code. Whereas the Color code possesses geometric features and symmetries, the Golay code exhibits algebraic ones. Considering that prior work incorporated geometric properties of the Toric code directly into the design of the continuous function, it is reasonable to infer that the effect and reproducibility of re-optimization may depend on how well the continuous function—and the multilayer perceptron approximating it can capture the structural characteristics of the code. Previous studies have reported that multilayer perceptrons tend to be more effective at learning geometric structures than algebraic ones, and the results of our experiments may have been influenced by this tendency \cite{amari2001}. Indeed, the structural evaluation of the continuous function approximation confirms that the geometric advantage of the color code is well reproduced. This suggests that, in the design of generalized continuous functions, the multilayer perceptron exhibits an inherent advantage in learning geometric structures.

\section{Conclusions} 
\label{4}
We proposed a re-optimization method for deep-learning–based decoders that does not rely on the specific construction of the stabilizer generators, thereby addressing the non-uniqueness of error patterns corresponding to syndrome measurement outcomes in stabilizer codes. 

We re-optimized Transformer decoders trained on the decoding problems of both the Color code and Golay code and observed improvements in decoding accuracy for both codes. For the case of $p = 0.05$, the improvements were approximately 0.8\% for the Color code and 0.1\% for the Golay code. A plausible explanation for why the improvement was larger for the Color code is that multilayer perceptrons (MLPs) are generally more effective at learning geometric structures than algebraic ones. In prior work on the Toric code, the geometric structure of the code was directly incorporated into the design of the continuous function, resulting in a larger accuracy gain than in our study. These observations suggest that the magnitude of the improvement depends not only on whether an MLP can approximate a similar function, but also on whether it can adequately capture the underlying structural properties of the original function. The results of the structural evaluation of the continuous function approximation also reveal differences in the reproduction of geometric structures, indicating that incorporating the structural properties of the code into function design and that approximation can be an effective strategy. Because we used an MLP to approximate the continuous function, better results were obtained for the Color code, which has geometric structure. However, adopting a deep-learning architecture capable of effectively learning algebraic structures may lead to greater improvements for the Golay code as well.

\bibliographystyle{unsrt}  
\bibliography{reference}    

\end{document}